\documentclass[twoside,11pt]{article}
\usepackage{amsfonts}
\usepackage{amsmath}
\catcode`@=11  %to allow use of @ as a letter; reset at the end
\font\fnotefont=cmr10 scaled \magstephalf
\def\fnote#1#2{\let\@sf\empty % parameter #2 (the text) is read later
  \ifhmode\edef\@sf{\spacefactor=\the\spacefactor}\/\fi
  {$\,^{#1}$}\@sf\vfootnote{#1}{#2}}
\def\vfootnote#1#2{\insert\footins\bgroup
  \interlinepenalty\interfootnotelinepenalty
  \splittopskip=\ht\strutbox % top baseline for broken footnotes
  \splitmaxdepth=\dp\strutbox \floatingpenalty=20000
  \ifdim\lastskip<2truept \removelastskip\vskip 2truept\fi
  \leftskip=\parindent \rightskip=0pt \spaceskip=0pt \xspaceskip=0pt
  \baselineskip=11pt\parindent=0pt\fnotefont
  \textindent{$\,^{#1}$}#2 \footstrut\vskip 3truept \egroup}
\def\footstrut{\vbox to\splittopskip{}}

\def\head#1 {\begin{center} {\bf #1}
              \end{center} \vskip 10pt}
\def\authors#1 {\begin{center} {\bf #1} \end{center}}
\def\sect #1 \par{\medbreak\vskip 4pt
     \begin{center} {\bf #1} \end{center} \par} \vskip 2pt
\long\outer\def\proposition#1 #2 \par{\smallbreak\vskip 4pt
    {{\sc Proposition #1.}\enspace {\sl #2}}\par
}
\long\outer\def\theorem#1 #2 \par{\smallbreak\vskip 4pt
    {{\sc Theorem #1.}\enspace {\sl #2}}\par
}
\long\outer\def\lemma#1 #2 \par{\smallbreak\vskip 4pt
    {{\sc Lemma #1.}\enspace {\sl #2}}\par
}
\long\outer\def\corollary#1 #2 \par{\smallbreak\vskip 4pt
    {{\sc Corollary #1.}\enspace {\sl #2}}\par
}
\outer\def\proof{\smallbreak\vskip 4pt {P r o o f.}\enspace}
\def\bibname{\bf References} % <--- NEW DEFINITION !!!

\def\thebibliography#1{\hfil \bibname \hfill \list
{[\arabic{enumi}]}{\settowidth\labelwidth{[#1]}\leftmargin
\labelwidth  \advance\leftmargin\labelsep
\usecounter{enumi}}
\def\newblock{\hskip .11em plus .33em minus .07em}
\sloppy\clubpenalty4000\widowpenalty4000
\sfcode`\.=1000\relax}

\newcommand{\bs}{Black-Scholes\ }

\newcommand{\half}{\frac{1}{2}}
\newcommand{\eqnref}[1]{Eq.~(\ref{eqn:#1})}

\newcommand{\hfunc}[4]{H{}_{#1}^{#2}{}_{#3}^{#4}}

\begin{document}
\setcounter{page}{1}
\thispagestyle{empty}
%%%%%%% making running heads on even and odd pages %%%%%%%
\markboth
% put in {} below the Author(s) without numberings
% (if many \authors) and with Initials for first name(s) only
{\rm \centerline{Michelle M. Wyss, Walter Wyss}}
% put in {} below a RUNNING HEAD (SHORT TITLE OF PAPER)
{\rm \centerline{EVOLUTION, ITS FRACTIONAL EXTENSION AND GENERALIZATION}}
%%%%%%%%%%%%%%%%%%%%%%%%%%%%%%%%%%%%%%%%%%%%%%%%%%%%%%%%%%
\vbox to 3.5cm {   % gives some real \vspace 4.5 cm !
\vfill
}  %%% to make empty space for Journal's Heading
%%%%%%%%%%%%% BODY OF PAPER %%%%%%%%%%%%%%%%%%%%%%%%%%%%%%
% HEAD
% put in {} THE TITLE OF THE PAPER in Capitals,
% if more than one line, use \\[6pt]

\head{EVOLUTION, ITS FRACTIONAL EXTENSION AND GENERALIZATION}  %%% or:  \head{ ... \\[6pt] ...}

% Authors
% put in {} The full name of the author(s), use
% numbering if more than one author

\authors{Michelle M. Wyss \&
         Walter Wyss$^*$}   % or:  \authors{..... \& .....$^{**}$}

% Footnote, to acknowledge grants and support
% fill in the second {} below with "Supported by ..."

%\fnote {} {}      %%% use like this
%\fnote {} {$^*$ Partially supported by ...}

\sect{Abstract}

%%% Text of obligatory abstract follows in 5-10 lines

The evolution of a quantity, described by a function of space and time, relates
the first derivative in time of this function to a spatial operator applied to
the function.  The initial value of the function at time $t=0$ is given.

The fractional extension of this evolution consists of replacing the first
derivative in time by a fractional derivative of order $\alpha$, 
$0 < \alpha \le 1$.

We give a relationship between the solution of the equation of evolution and the
solution of the equation belonging to its fractional extension.

\vskip 6pt

{\it Mathematics Subject Classification}: 26A33

{\it Key Words and Phrases}: equation of evolution,  fractional
calculus, fractional diffusion, fractional \bs equation, 
Fox's $H$-functions, Generalized Mittag-Leffler functions.

\vskip 3pt

\sect{1. Introduction}

Let $u(x,t)$ be a function of space and time; $x \in \Re^n$, $t \in \Re_+$. The
equation, with initial condition,
\begin{align}
\frac{\partial u(x,t)}{\partial t} & = \left[L(x) u \right] (x,t) \label{eqn:w1}\\
u(x,0) & = f(x) \label{eqn:w2}
\end{align}
is called an equation of evolution.  The linear spatial operator $L(x)$ depends on the
problem at hand.  We assume that the solution $u(x,t)$ of this problem has been
found.

The fractional extension of the above equation of evolution is given by the
integral equation
\begin{gather}
u_{\alpha}(x,t) = u(x,0) + \frac{1}{\Gamma(\alpha)} 
\int_0^t d\tau\, (t-\tau)^{\alpha-1} \left[L(x) u_{\alpha} \right](x,\tau) \\
0<\alpha \le 1 \, .
\end{gather}
For $\alpha = 1$ we recover the solution of the equation of evolution. 

In this paper we find a relationship between the solutions $u_{\alpha}(x,t)$
and $u(x,t)$. A generalization of this relationship is also proposed.

\sect{2. Evolution and its fractional extension I}

The equation of evolution Eqs.~(\ref{eqn:w1}),~(\ref{eqn:w2}) can be written as 
\begin{align}
u(x,t) &= f(x) + \int_0^t d\tau\, \left[L(x) u \right](x,\tau) \label{eqn:w5} \\
u(x,0) &= f(x)  \, .
\end{align}
The Laplace transform $\tilde{u}(x,p)$ in the variable $t$ is given by
\begin{align}
\tilde{u}(x,p) &= \int_0^{\infty} dt\, e^{-pt} u(x,t) \\
\intertext{and leads to} 
\left[L(x) \tilde{u} \right](x,p) &= p \tilde{u}(x,p) -f(x) \, .
\label{eqn:w8}
\end{align}
The fractional extension of the above equation of evolution reads
\begin{equation}
u_{\alpha}(x,t) =f(x)+ \frac{1}{\Gamma(\alpha)} \int_0^t d\tau\, (t-\tau)^{\alpha-1} \left[L(x) u_{\alpha} \right](x,\tau) \, .
\label{eqn:w9}
\end{equation}

The Laplace transform $\tilde{u}_{\alpha}(x,p)$ in the variable $t$ is given by
\begin{align}
\tilde{u}_{\alpha}(x,p) &= \int_0^{\infty} dt\, e^{-pt} u_{\alpha}(x,t) \\
\intertext{and leads to} 
\left[L(x) \tilde{u}_{\alpha} \right](x,p) &= p^{\alpha} \tilde{u}_{\alpha}(x,p)
-f(x)p^{\alpha-1} \, .
\label{eqn:w11}
\end{align}

\lemma{1}
The solutions 
$\tilde{u}(x,p)$ and $\tilde{u}_{\alpha}(x,p)$ are related by
\begin{equation}
\tilde{u}_{\alpha}(x,p) = p^{\alpha-1} \tilde{u} \left( x,p^{\alpha} \right) \, .
\label{eqn:w12}
\end{equation}

\proof 
\begin{align}
\left[ L(x)\tilde{u}_{\alpha} \right] (x,p) &= p^{\alpha-1}\left[ L(x)\tilde{u}\right]
\left( x,p^{\alpha} \right) \\
   &= p^{\alpha-1}\left\{p^{\alpha} \tilde{u}\left( x,p^{\alpha} \right) - f(x)
      \right\} \\
   &= p^{\alpha}\tilde{u}_{\alpha}(x,p)- f(x)p^{\alpha-1}
\end{align}
which is \eqnref{w11}.

We now look at the Mellin transforms of $u(x,t)$ and $u_{\alpha}(x,t)$ in the
variable $t$.
\begin{align}
\hat{u}(x,s) &= \int_0^\infty dt\, t^{s-1} u(x,t) \\
\hat{u}_{\alpha}(x,s) &= \int_0^\infty dt\, t^{s-1} u_{\alpha}(x,t)   \, .
\end{align}

The relationship between the Mellin and Laplace transform of a function
$\phi(t)$ is
\begin{equation}
\hat{\phi}(s) = \frac{1}{\Gamma(1-s)} \int_0^\infty dp\, p^{-s} \tilde{\phi}(p) \, .
\label{eqn:w18}
\end{equation}

\lemma{2}
$\hat{u}(x,s)$ and $\hat{u}_\alpha(x,s)$ are related by 
\begin{equation}
\hat{u}_{\alpha} (x,s) = \frac{1}{\alpha} 
\frac{\Gamma\left(1-\frac{s}{\alpha}\right)}{\Gamma(1-s)}
\hat{u}\left(x,\frac{s}{\alpha}\right) \, .
\label{eqn:w19}
\end{equation}

\proof
From Eqs. (\ref{eqn:w18}), (\ref{eqn:w12}) we get 
\begin{equation}
\hat{u}_\alpha(x,s) = \frac{1}{\Gamma(1-s)} \int_0^\infty dp\, p^{-s} p^{\alpha -1}
\tilde{u}\left( x,p^{\alpha} \right) \, .
\end{equation}
The change of variable 
\begin{equation}
p^{\alpha} = q
\end{equation}
leads to
\begin{align}
\hat{u}_\alpha(x,s) &= \frac{1}{\alpha}\frac{1}{\Gamma(1-s)} \int_0^\infty dq\,
q^{-\frac{s}{\alpha}} \tilde{u}\left( x,q \right) \\
\intertext{or}
\hat{u}_\alpha(x,s) &= \frac{1}{\alpha}\frac{1}{\Gamma(1-s)} 
\Gamma\left(1-\frac{s}{\alpha}\right) \hat{u}\left(x,\frac{s}{\alpha}\right) \, .
\end{align}
This is \eqnref{w19}.

\sect{3. Generalized Mittag-Leffler Functions}

The generalized  
Mittag-Leffler
functions~\cite{Schneider} are given by 
\begin{equation}
F_{\alpha\beta}(z) = \Gamma(\beta) \sum_{k=0}^{\infty} (-1)^k 
\frac{1}{\Gamma(\beta + \alpha k)} z^k 
\label{eqn:w24}
\end{equation}
with 
\begin{equation}
z \ge 0 \; , \; 0 < \alpha \le 1 \; , \; \beta \ge \alpha \, .
\end{equation}
$\beta = 1$ give the usual Mittag-Leffler functions which we denote by
$F_\alpha(z)$.

The $H$-function representation of
$F_{\alpha\beta}(z)$~\cite{Fox,Braaksma,Srivastava} is given by
\begin{equation}
F_{\alpha\beta}(z) = \Gamma(\beta) \hfunc{1}{1}{2}{1} 
\left( z \biggl| 
\begin{matrix}
(0,1) \\
(0,1)(1-\beta,\alpha)
\end{matrix}
\right) \, .
\end{equation}
The generalized Mittag-Leffler functions are the Laplace transform of probability
measures on $\Re_+$.  The corresponding probability densisties are given by
\begin{equation}
f_{\alpha\beta}(z) = \Gamma(\beta) \hfunc{1}{1}{1}{0} 
\left( z \biggl| 
\begin{matrix}
(\beta-\alpha,\alpha) \\
(0,1)
\end{matrix}
\right) \, .
\label{eqn:w27}
\end{equation}
$\beta = 1$ gives the probability density of the usual Mittag-Leffler functions
which is denoted by $f_\alpha (z)$.  It has the $H$-function representation
\begin{align}
f_{\alpha}(z) &=  \hfunc{1}{1}{1}{0} 
\left( z \biggl| 
\begin{matrix}
(1-\alpha,\alpha) \\
(0,1)
\end{matrix}
\right)
\; , \; 0<\alpha<1 \label{eqn:w28} \\
\intertext{and}
f_1(z) &= \delta(z-1) \, .
\end{align}

For $0<\alpha<1$, $f_\alpha(z)$ is an entire function and vanishes exponentially
for large positive $z$.  It has the power series representation
\begin{gather}
f_\alpha(z) = \sum_{k=0}^\infty (-1)^k \frac{1}{\Gamma(1-\alpha-\alpha k)}
\frac{z^k}{k!} \\
0<\alpha<1 \; , \; z \in \Re_+ \, .
\end{gather}

\sect{4. Evolution and its Fractional Extension II}

The Mellin transform of the probability density $f_\alpha (z)$ is given through
\eqnref{w28} as
\begin{equation}
\hat{f}_\alpha (s) = \frac{\Gamma(s)}{\Gamma(1-\alpha+\alpha s)} \, .
\label{eqn:w32}
\end{equation}

\lemma{3}
The inverse Mellin transforms of $\hat{u}(x,s)$ and $\hat{u}_\alpha(x,s)$ are
related by 
\begin{equation}
u_{\alpha} (x,t) = \int_0^\infty dz\, f_\alpha(z) 
u \left( x,t^\alpha z \right) \, .
\label{eqn:w33}
\end{equation}

\proof
Let  $M$ denote the performance of the Mellin transform.  From~\cite{Erdely} we
have
\begin{align}
\left[ M \int_0^\infty dz\, f(tz) g(z) \right] (s) &= \hat{f}(s)\hat{g}(1-s) \\
\left[ M f(at^\alpha) \right] (s) &= \frac{1}{\alpha} a^{-\frac{s}{\alpha}} 
\hat{f}\left(\frac{s}{\alpha}\right) \\
\left[ M \int_0^\infty dz\, f(z) g(t^\alpha z) \right] (s) &= \frac{1}{\alpha}
\left[ M \int_0^\infty dz\, f(z) g(tz) \right] \left(\frac{s}{\alpha}\right) \, .
\end{align}

From \eqnref{w32} we find
\begin{equation}
\hat{f}_\alpha \left(1-\frac{s}{\alpha}\right) = 
\frac{\Gamma \left (1-\frac{s}{\alpha}\right)}{\Gamma(1-s)}
\end{equation}
and thus from \eqnref{w19}
\begin{equation}
\hat{u}_\alpha (x,s) = \frac{1}{\alpha} \hat{f}_\alpha \left(1-\frac{s}{\alpha}\right)
\hat{u}\left(x,\frac{s}{\alpha}\right) \, .
\end{equation}
This leads to
\begin{align}
\hat{u}_\alpha (x,s) &= \frac{1}{\alpha} \left[ M \int_0^\infty dz\, f_\alpha(z)
                       u(x,t z) \right] \left(\frac{s}{\alpha}\right) \\
                     &= \left[ M \int_0^\infty dz\, f_\alpha(z)
                        u(x,t^\alpha z) \right] (s)  \, .
\end{align}
Thus
\begin{equation}
u_\alpha (x,t) = \int_0^\infty dz\, f_\alpha(z) u(x,t^\alpha z)  \, .
\end{equation}

\theorem{1} The solution $u(x,t)$ of the equation of evolution \eqnref{w5} and
the solution $u_\alpha(x,t)$ of its fractional extension \eqnref{w9} are related
by
\begin{equation}
u_\alpha (x,t) = t ^{-\alpha} \int_0^\infty dz\, 
f_\alpha \left( t^{-\alpha} z \right) u(x,z)  \, .
\label{eqn:w42}
\end{equation}

\proof
Make the variable transform $t^\alpha z = w$ in \eqnref{w33}.

\sect{5. The Green's functions}

The solution of the equation of evolution \eqnref{w5} can be written as
\begin{equation}
u(x,t) = \int dy\, G(x,y;t) f(y)
\end{equation}
where we assume that the Green's function $G(x,y;t)$ is known.

The solution of the equation of its fractional extension \eqnref{w9} can be
written as 
\begin{equation}
u_\alpha (x,t) = \int dy\,  G_\alpha(x,y;t) f(y) \, .
\end{equation}
According to Theorem~1, \eqnref{w42} the two Green's functions are related by
\begin{equation}
G_\alpha (x,y;t) = t^{-\alpha} \int_0^\infty dz\, 
f_\alpha \left( t^{-\alpha} z \right) G(x,y;t) \, .
\end{equation}

\sect{6. Applications}

\begin{enumerate}
\item{
The fractional diffusion equation~\cite{SchneiderWyss}. The equation of evolution is
given by
\begin{equation}
\frac{\partial u(x,t)}{\partial t} = \Delta u(x,t)
\label{eqn:w46}
\end{equation}
where 
\begin{equation}
\Delta = \sum_{k=1}^{n} \frac{\partial^2}{\partial x_k^2}
\end{equation}
and the initial condition
\begin{equation}
u(x,0) = f(x) \, .
\end{equation}

The Green's function for this problem is 
\begin{equation}
G(x,y;t) = G(|x-y|,t)
\end{equation}
where with 
\begin{align}
r &=  |x-y| \\
G(r,t) &= (4 \pi t) ^{-\frac{n}{2}} e^{-\frac{r^2}{4t}} \, .
\end{align}
The Green's function for the fractional extension of \eqnref{w46} is then
\begin{equation}
G_\alpha(r,t) = t^{-\alpha} \int_0^\infty dz\, f_\alpha \left(t^{-\alpha} z
\right) G(r,z) \, .
\end{equation}
We compute this integral by looking at its Mellin transform in the variable $t$.
\begin{align}
\hat{G}_\alpha (r,s) &= \frac{1}{\alpha} 
\hat{f}_\alpha \left( 1 - \frac{s}{\alpha} \right)
\hat{G} \left( r,\frac{s}{\alpha} \right) \\
&= \frac{1}{\alpha} \frac{\Gamma\left( 1 - \frac{s}{\alpha} \right)}{\Gamma(1-s)}
\hat{G} \left( r,\frac{s}{\alpha} \right)  \, .
\end{align}

From~\cite{Erdely} we get
\begin{equation}
\hat{G} (r,s) =  \pi^{-\frac{n}{2}} 2^{-2s} r^{2s-n} 
\Gamma \left(\frac{n}{2} - s \right)
\end{equation}
and thus
\begin{equation}
\hat{G}_\alpha (r,s) = \frac{1}{\alpha} \pi^{-\frac{n}{2}} 
2^{-\frac{2s}{\alpha}} r^{\frac{2s}{\alpha}-n} 
\Gamma \left(\frac{n}{2} - \frac{s}{\alpha} \right)
\frac{\Gamma\left( 1 - \frac{s}{\alpha} \right)}{\Gamma(1-s)} \, .
\end{equation}

Inverting the Mellin transform leads to
\begin{equation}
G_\alpha (r,t) = \pi ^{-\frac{n}{2}} 2^{-1} r^{-n} 
\hfunc{1}{2}{2}{0} \left( \half r t^{-\frac{\alpha}{2}} \biggl| 
\begin{matrix}
(1,\frac{\alpha}{2}) \\
(\frac{n}{2}, \half) (1, \half)
\end{matrix}
\right) \, .
\end{equation}
This is indeed the Green's function as given in~\cite{SchneiderWyss}.
}

\item{Fractional Black-Scholes equation~\cite{Wyss:fbse}

The equation of evolution (Black-Scholes equation) is given by
\begin{gather}
\frac{\partial C}{\partial t} + \half \sigma^2 S^2 
\frac{\partial^2 C}{\partial S^2} + rS \frac{\partial C}{\partial S} - rC = 0 \\
S \ge 0 \; , \; t \le T
\end{gather}
and boundary condition
\begin{equation}
C(S,T) = \max \{ S-E,0 \}  \, .
\end{equation}
With the transformation
\begin{align}
t         &= T - \frac{2}{\sigma^2} \tau \; , \; C(S,t) = A(S,\tau) \\
\lambda_0 &= \frac{2r}{\sigma^2} 
\end{align}
we have the equation of evolution
\begin{equation}
\frac{\partial A}{\partial \tau} = S^2 \frac{\partial^2 A}{\partial S^2} +
\lambda_0 S \frac{\partial A}{\partial S} - \lambda_0 A
\end{equation}
with the initial value
\begin{equation}
A(S,0) = \max \{ S-E,0 \}  \, .
\end{equation}

The solution is given by
\begin{equation}
A(S,\tau) = S N(d_1) - E e^{-\lambda_0 \tau} N(d_2)
\end{equation}
where 
\begin{align}
d_1 &= (2\tau)^{-\half} \left[ \ln \left( \frac{S}{E} \right)  + (\lambda_0 + 1) \tau \right] \\
d_2 &= (2\tau)^{-\half} \left[ \ln \left( \frac{S}{E} \right)  + (\lambda_0 - 1) \tau \right] 
\end{align}
and 
\begin{equation}
N(d) = \frac{1}{\sqrt{2\pi}} \int_{-\infty}^{d} dx\, e^{-\half x^2} \, .
\end{equation}

According to Theorem 1 \eqnref{w42} the solution of the fractional extension of the Black-Scholes equation
\begin{multline}
A_\alpha (S,\tau) = A(S,0) + \frac{1}{\Gamma(\alpha)} \int_0^\tau dz\, (\tau - z)^{\alpha -1} 
\\ \times \left[ S^2 \frac{\partial^2 A_\alpha (S,z)}{\partial S^2} 
+ \lambda_0 S  \frac{\partial A_\alpha (S,z)}{\partial S} - \lambda_0 A_\alpha (S,z) \right]
\end{multline}
is given by
\begin{equation}
A_\alpha (S,\tau) = \tau^{-\alpha} \int_0^\infty dz\, f_\alpha \left( \tau^{-\alpha} z \right) A(S,z) \, .
\end{equation}
}
\end{enumerate}

\sect{7. Generalization}

According to \eqnref{w33} the solution of the equation for the fractional extension of an
equation of evolution is given by \eqnref{w33}
\begin{equation}
u_\alpha (x,t) = \int_0^\infty dz\, f_\alpha(z) u \left( x,t^\alpha z \right)
\label{eqn:w70}
\end{equation}
$f_\alpha(z)$ is a measure on $\Re_+$ and $f_1(z) = \delta(z-1)$.

The Laplace transform of $f_\alpha$ is the usual Mittag-Leffler function.

The generalized Mittag-Leffler function \eqnref{w24} is the Laplace transform of the measure
$f_{\alpha\beta}(z)$ on $\Re_+$, \eqnref{w27}.  We have
\begin{align}
f_{\alpha\beta}(x) &= \Gamma(\beta) \sum_{k=0}^{\infty} (-1)^k 
\frac{1}{\Gamma(\beta - \alpha - \alpha k)} \frac{x^k}{k!} 
\; , \; a<1 \; , \; \beta \ge \alpha \\
f_{1 \beta}(x) &= \left\{  \begin{matrix}
(\beta - 1)(1-x)^{\beta-2} & , & 0 \le x \le 1  \\
0 & , & 1<x
\end{matrix} \right.
\; , \; \beta>1
\\
f_{11} (x) &= \delta(x-1) \, .
\end{align}

It is thus tempting to generalize from the solution $u_\alpha (x,t)$ in \eqnref{w70} to 
\begin{equation}
u_{\alpha \beta} (x,t) = \int_0^\infty dz\, f_{\alpha \beta} (z) 
u_{\alpha \beta} (x,t^\alpha z) \, .
\end{equation}
This expression uses the form factor $f_{\alpha\beta}$ and will be studied elsewhere.  Other
form factors might also be considered.

\sect{8. Summary}

We assume that we know the solution of a general equation of evolution.  The
solution of the equation of its fractional extension is then related to the
above solution by what we call a form factor.  This form factor belongs to the
Mittag-Leffler function. A fractional generalization is proposed by using the
form factor belonging to the generalized Mittag-Leffler function.

\leftskip 2pc \parindent -2pc

\vskip 1cm
\bibliographystyle{unsrt}
\bibliography{wyss}
\vskip 1cm

\it

$^{*}$) Dept. of Physics \hfill  Received: Month,Date,Year

University of Colorado

Boulder, CO 80309

e-mail: w2ch@hotmail.com

\end{document}